\documentclass[final,3p,times]{elsarticle}

\usepackage{amsmath, amssymb}
\usepackage{amsthm}
\usepackage{natbib}
\usepackage{soul}
\usepackage[colorlinks,citecolor=blue, urlcolor=blue, linkcolor=blue]{hyperref}
\usepackage[normalem]{ulem}               
\newcommand\redout{\bgroup\markoverwith
{\textcolor{red}{\rule[0.5ex]{2pt}{0.8pt}}}\ULon}
\newcommand\blueout{\bgroup\markoverwith
{\textcolor{blue}{\rule[0.5ex]{2pt}{0.8pt}}}\ULon}

\journal{Computer Physics Communications}

\newcommand{\dn}{\downarrow}
\newcommand{\up}{\uparrow}

\begin{document}

\begin{frontmatter}

 \title{The {\it victory} project v1.0: an efficient parquet equations solver }

\author[SHT,WIEN]{Gang Li}
\author[WIEN]{Anna Kauch}
\author[WIEN,SHT]{Petra Pudleiner}
\author[WIEN]{Karsten Held}
\address[SHT]{School of Physical Science and Technology, ShanghaiTech University, Shanghai 200031, China}
\address[WIEN]{Institute of Solid State Physics, Vienna University of Technology, A-1040 Vienna, Austria}

\begin{abstract} 
{\it Victory}, {\it i.e.} \underline{vi}enna \underline{c}omputational \underline{to}ol deposito\underline{ry}, is a collection of numerical tools
for solving the parquet equations for the Hubbard model and similar many body problems. The  parquet formalism is
a self-consistent theory at both the single- and two-particle levels, and  can thus describe individual fermions as well as their collective behavior on equal footing. This is essential for the understanding of various emergent phases and their transitions in many-body systems, in particular for cases in which a single-particle description fails.
Our implementation of {\it victory} is in modern Fortran and it fully respects the structure of various vertex functions in both momentum and Matsubara frequency space. 
We found the latter to be crucial for the convergence of the parquet equations, as well as for the correct determination of various physical observables. 
In this release, we thoroughly explain the program structure and the controlled approximations to efficiently solve the parquet equations, {\it i.e.} the two-level kernel approximation and the high-frequency regulation. 
\begin{flushleft}
\textbf{Program summary}
\item {\it Program title:} Victory
\item {\it Programming language:} Fortran 90
\item {\it Parallel mode:} Segments of vertex functions are distributed over different cores. MPI{\_}BCAST is adopted for efficient data exchange. 
\item {\it Operating system:} Unix, Linux, Windows
\item {\it Version history:} 1.0
\item {\it Nature of the problem:} 
The parquet equations require the knowledge of the fully irreducible vertex 
from which all one- and two-particle vertex and Green's functions are calculated. The underlying two-particle vertex functions are large memory objects that depend on three  momenta with periodic  boundary conditions and three frequencies with open ones. The coupled diagrams of the parquet equations extend the frequency dependence of the reducible vertex functions to a larger frequency space where, {\em a priori}, no information is available. 
\item {\it Solution method:} The reducible vertex functions are found to possess the simplest structure among all two-particle vertex functions and can be approximated  by functions with a reduced number of arguments, {\it i.e.} the kernel functions. The open boundary issue of the vertex functions in Matsubara-frequency space is then solved as follows: we solve the parquet equations with the reducible vertex functions whenever it is possible, otherwise we supplement these by the kernel functions.   
\end{flushleft}
\end{abstract}


\end{frontmatter}

\section{Introduction}
\label{Sec:introduction}
The understanding of materials with strong electronic correlations, such as transition metal oxides and $f$-electron systems, represents one of the greatest challenges to contemporary condensed-matter theory and remains a subject of intense research. 
The localized $d$- or $f$-electrons are subject to strong Coulomb interactions among themselves, which in turn causes a breakdown of a single-particle description in these systems.    
The problem arises essentially due to the presence of different energy scales which compete with each other in many-body systems~\cite{RevModPhys.84.299,Martin16}, and which is in fact  the source of many intriguing phenomena.   
The model describing the competition of the electronic kinetic energy $\epsilon_{\mathbf{k}}$ and the Coulomb repulsion $U$ is the so-called Hubbard model
\begin{equation}\label{Hubbard_Model}
H = \sum_{\mathbf{k},\sigma}\epsilon_{\mathbf{k}}c_{\mathbf{k}\sigma}^{\dagger}c_{\mathbf{k}\sigma}^{\phantom{\dagger}}+U\sum_{i}n_{i\uparrow}n_{i\downarrow}\;.
\end{equation}
Here,  ${c}^{({\dagger})}_{\mathbf{k} \sigma}$ annihilates (creates) an electron with momentum ${\mathbf{k}}$; and $n_{i\sigma}\equiv c_{i\sigma}^{\dagger}c_{i\sigma}^{\phantom{\dagger}}$ is the occupation number operator at lattice site $i$.
The two terms in the Hubbard model describe single- and two-particle processes, respectively. 
The interacting electrons repel each other with interaction $U$, and hence
 tend to avoid each other to minimize the energy.
As a result the charge carriers loose their mobility and become localized in a Mott insulator.  
As a competing effect, the kinetic energy  $\epsilon_{\mathbf{k}}$ is minimized by increasing the itinerancy of the electrons, which drives the system away from a localized state. 

In addition to the two apparent energy scales, more competing energy scales can emerge in this model. For example, when the system temperature becomes lower than a certain condensation energy, the entire many-body system will behave collectively as composed particles.
A well-known example for this is the high-$T_{c}$ superconductor, where the composite electronic degrees of freedom are the Cooper pairs.
Similarly, the condensation of particle-hole pairs can give rise to charge (CDW) or spin density wave (SDW) orders depending on the broken symmetry, respectively. 
Very often a many-body system displays both, the single-particle and the composed-particle characters, as well as their phase transitions. Here, an understanding of the single- {\em and} two-particle response is required on an equal footing.
 
 This is what the parquet equations~\cite{Dominicis_Martin_1, Dominicis_Martin_2} that are  implemented in the {\it victory} package provide.
The parquet formulation for condensed matter physics has been discussed thoroughly in~\cite{doi:10.1142/S021797929100016X,PhysRevB.55.2122,PhysRevB.43.8044,PhysRevLett.62.961,Bickers-Review} and is one of the very few theories that maintains self-consistency at both single- and two-particle level.
It requires knowing the two-particle fully irreducible vertex, or approximating it, e.g. by the bare interaction $U$ in the parquet approximation \cite{footnote2}, by a self-consistently obtained constant \cite{Janis2017} or by the local vertex of an  impurity problem as in diagrammatic extensions of dynamical mean field theory  \cite{Toschi2007,doi:10.1143/PTPS.176.117,Rubtsov2008, PhysRevLett.102.206401, Rohringer2013,Taranto2014,Ayral2015,Li2015,RMPVertex}.
In the present paper, we will discuss the detailed implementation of the parquet equations in the  {\it victory} package.

The outline of the paper is as follows:
In Section
\ref{Sec:program}, we describe the program structure.
Here, we start by recapitulating the parquet equations in Section \ref{Sec:parquet} and discuss the employed symmetries in Section
\ref{Sec:symmetry}.  The structure of the {\it victory} program is outlined  in Section \ref{Sec:programstructure}, important technical aspects are discussed in Section 
\ref{Sec:technical}, and the parallelization strategy in Section
\ref{Sec:parallelization}.
In Section
\ref{Sec:example} we discuss ---as an example--- the application to the Hubbard model on a square lattice and present one- and two-particle results in Sections
\ref{Sec:singleparticle} and
\ref{Sec:twoparticle}, respectively. Finally, Section
\ref{Sec:conclusion} provides a brief conclusion and outlook.

\section{Description of the program structure}
\label{Sec:program}
\subsection{Parquet equations}
\label{Sec:parquet}
The parquet formalism consists of a set of coupled equations, {\it i.e.} the Bethe-Salpeter, the parquet and the Schwinger-Dyson equations, which include both single-particle Green's function and two-particle vertex functions. 
To be self-contained and at the same time to keep the discussion simple,  let us first introduce the necessary functions which contain all topologically invariant one- and two- particle Feynman diagrams.
Let us start with the one-particle functions: The 
 self-energy $\Sigma$ contains all irreducible \cite{footnote1} diagrams with one incoming and one outgoing particle (leg). From it all (reducible) Green's function $G$ diagrams are obtained through 
 the Dyson equation 
\begin{equation}\label{GF}
G(k) = [G^{-1}_{0}(k) - \Sigma(k)]^{-1} = \left[ i\nu + \mu - \epsilon_{\mathbf{k}} - \Sigma(k)\right]^{-1}\;,
\end{equation}
with the chemical potential $\mu$, the fermionic Matsubara frequency $\nu$, and a  combined notation of momentum $\mathbf{k}$ and frequency $\nu$, {\em i.e.} $k=(\mathbf{k}, \nu)$. The non-interacting dispersion relation $\epsilon_{\mathbf{k}}$ directly yields the non-interacting Green's function
$G_{0}(k)=( i\nu + \mu - \epsilon_{\mathbf{k}} )^{-1}$.

Regarding two-particle quantities, the concept of reducibility is more subtle.
The full vertex functions $F$ can be further subdivided into different classes, namely a fully irreducible class of diagrams $\Lambda$ and classes of diagrams $\Phi_r$ which are reducible regarding a specific channel $r$.
In the $\mbox{SU}(2)$-symmetric case, it is convenient to use some combinations of the  spin indices known as the density ($d$), magnetic ($m$), singlet ($s$) and triplet ($t$) channel (in addition to the   $r$ channels regarding the irreducibility).
This leads to the following parquet equations, see~\cite{Bickers-Review,RMPVertex, GangLi-1}:
\begin{subequations}\label{PA_F}
\begin{align}
F_{d}^{k,k^{\prime}}(q)=\Lambda_{d}^{k,k^{\prime}}(q) + \Phi^{k,k^{\prime}}_{d}(q) - \frac{1}{2}\Phi_{d}^{k,k+q}(k^{\prime}-k) - \frac{3}{2}\Phi_{m}^{k,k+q}(k^{\prime}-k)
+\frac{1}{2}\Phi_{s}^{k,k^{\prime}}(k+k^{\prime}+q)
+\frac{3}{2}\Phi_{t}^{k,k^{\prime}}(k+k^{\prime}+q)\;;\\
F_{m}^{k,k^{\prime}}(q)=\Lambda_{m}^{k,k^{\prime}}(q) + \Phi^{k,k^{\prime}}_{m}(q) - \frac{1}{2}\Phi_{d}^{k,k+q}(k^{\prime}-k)
+ \frac{1}{2}\Phi_{m}^{k,k+q}(k^{\prime}-k)
 -\frac{1}{2}\Phi_{s}^{k,k^{\prime}}(k+k^{\prime}+q)
+\frac{1}{2}\Phi_{t}^{k,k^{\prime}}(k+k^{\prime}+q)\;;\\
F_{s}^{k,k^{\prime}}(q)=\Lambda_{s}^{k,k^{\prime}}(q) + \Phi^{k,k^{\prime}}_{s}(q) 
 +\frac{1}{2}\Phi_{d}^{k,q-k^{\prime}}(k^{\prime}-k)
 -\frac{3}{2}\Phi_{m}^{k,q-k^{\prime}}(k^{\prime}-k)
 +\frac{1}{2}\Phi_{d}^{k,k^{\prime}}(q-k-k^{\prime})
 - \frac{3}{2}\Phi_{m}^{k,k^{\prime}}(q-k-k^{\prime})\:;\\
F_{t}^{k,k^{\prime}}(q) = \Lambda_{t}^{k,k^{\prime}}(q)  + \Phi^{k,k^{\prime}}_{t}(q)
 -\frac{1}{2}\Phi_{d}^{k,q-k^{\prime}}(k^{\prime}-k)
 -\frac{1}{2}\Phi_{m}^{k,q-k^{\prime}}(k^{\prime}-k)
+\frac{1}{2}\Phi_{d}^{k,k^{\prime}}(q-k-k^{\prime})
+ \frac{1}{2}\Phi_{m}^{k,k^{\prime}}(q-k-k^{\prime})\:. 
\end{align}
\end{subequations} 
Here, $F_{d/m/s/t}^{k,k^{\prime}}(q)$ are the complete vertices in the corresponding channels (combinations of spin indices). 
The equations (\ref{PA_F}) respect an important symmetry of the two-particle vertex, {\it i.e.} the crossing symmetry~\cite{Bickers-Review}. With the above formulations, this symmetry has been guaranteed to be satisfied~\cite{PhysRevE.87.013311}. Note that, in principle,  $F_{s/t}$ can be obtained from  $F_{d/m}$ at a different combination of frequencies and momenta. But this would take us out of the finite frequency box the vertex is stored in and break the crossing symmetry.

The vertex functions $\Phi_{d/m}^{k,k^{\prime}}(q)$ and $\Phi_{s/t}^{k,k^{\prime}}(q)$ in (\ref{PA_F})  denote the reducible vertex in the particle-hole ($d/m$) and particle-particle ($s/t$) channel, respectively.
They can be calculated from the corresponding complete vertex $F_{d/m/s/t}^{k,k^{\prime}}(q)$ and irreducible vertex functions $\Gamma^{k,k^{\prime}}_{d/m/s/t}(q) = F_{d/m/s/t}^{k,k^{\prime}}(q) - \Phi_{d/m/s/t}^{k,k^{\prime}}(q)$ through the Bethe-Salpeter equations: 
\begin{subequations}\label{PA_F_Phi}
\begin{align}
\Phi_{d/m}^{k,k^{\prime}}(q) &= \frac{T}{N}\sum_{k^{\prime\prime}}\Gamma^{k,k^{\prime\prime}}_{d/m}(q)G(k^{\prime\prime})G(k^{\prime\prime}+q)F^{k^{\prime\prime},k^{\prime}}_{d/m}(q)\;,\\
\Phi_{t/s}^{k,k^{\prime}}(q) &= \pm\frac{T}{2N}\sum_{k^{\prime\prime}}\Gamma^{k,k^{\prime\prime}}_{t/s}(q)G(k^{\prime\prime})G(q-k^{\prime\prime})F^{k^{\prime\prime},k^{\prime}}_{t/s}(q)\;,
\end{align}
\end{subequations}
\begin{figure}[tb]
\centering
\includegraphics[width=.6\linewidth]{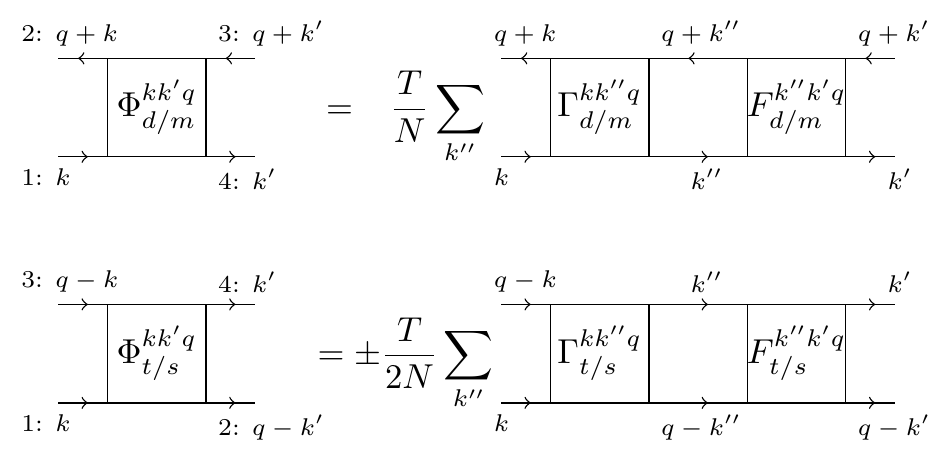}
\caption{The Bethe-Salpeter equations (\ref{PA_F_Phi})  in terms of Feynman diagrams. The notation corresponds to \cite{GangLi-1}.}
\label{fig:PA_F_Phi}
\end{figure}
where $T$ is temperature and $N$ is the number of momenta in the first Brillouin zone. To clarify the used notation, Fig.~(\ref{fig:PA_F_Phi}) explicitly illustrates Eq.~(\ref{PA_F_Phi}).

In the Bethe Salpeter Eqs.~(\ref{PA_F_Phi}), we need the one-particle Green's function as an input. For a self-consistent scheme, we hence need to recalculate the one-particle quantities from the two-particle vertices. This
is achieved via the (Heisenberg)  equation of motion, also known as Dyson-Schwinger equation: 
\begin{equation}\label{DysSchw}
\Sigma(k) = -\frac{UT^{2}}{2N^{2}}\sum_{k^{\prime},q}[F_{d}^{k,k^{\prime}}(q)-F_{m}^{k,k^{\prime}}(q)]G(k+q)G(k^{\prime})G(k^{\prime}+q)\;.
\end{equation}

\begin{figure}[tb]
\centering
\includegraphics[width=.55\linewidth]{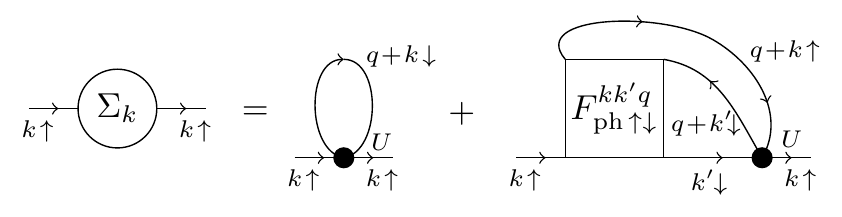}
\caption{These Feynman diagrams represent Eq.~(\ref{DysSchw}), where the identity $F_{\!ph\up\!\dn}^{k,k^{\prime}}(q)=F^{k,k^{\prime}}_d(q)-F^{k,k^{\prime}}_m(q)$ is used.}
\label{fig:DysSchw}
\end{figure}
Fig.~\ref{fig:DysSchw} shows the equivalent Feynman diagrams thereof.
At this stage, it is easy to see that given the fully irreducible vertex $\Lambda_{d/m/s/t}$ as input, the other two-particle vertices and the single-particle self-energy can be consistently determined from the parquet equations (\ref{PA_F}), the Bethe-Salpeter (\ref{PA_F_Phi}) and the Dyson-Schwinger equation (\ref{DysSchw}).
The Green's function is finally updated after each iteration via the Dyson Eq.~(\ref{GF}). This set of equations is solved self-consistently in {\em victory}.

\subsection{Symmetry consideration}
\label{Sec:symmetry} In addition to the particle-hole symmetry, the crossing symmetry and the spin $\mbox{SU}(2)$ symmetry, we also apply the time-reversal symmetry to the calculation of the parquet equation
in {\it victory}. 
For the two-particle vertex functions, taking $F^{k,k^{\prime}}_{r}(q)$ as an example, the time-reversal symmetry implies
\begin{equation}
\left[ F^{k,k^{\prime}}_{r}(q) \right]^{*} = F^{-k,-k^{\prime}}_{r}(-q)\;,
\end{equation}
which indicates that only the positive (negative) frequencies in either $k$, $k^{\prime}$ or $q$ are needed.
The negative (positive) counterparts can be obtained straightforwardly from the above time-reversal symmetry operation. 
In {\it victory}, we hence only define zero and positive frequencies for $q$. But for $k$ and $k^{\prime}$ the frequencies still need to be stored  from negative to positive as usual. 
Introducing the symmetry operation to the two-particle vertex has a clear advantage, as it can significantly reduce the memory requirement for storing the vertices which is the bottleneck for solving the parquet equations. This outweighs 
the  additional CPU operations needed for employing the symmetry.
Other symmetries that connect different momentum and frequency sectors can be straightforwardly implemented in {\it victory}, which can further reduce the size of the two-particle vertex functions. This is currently work in progress.

\subsection{Outline of the program structure}
\label{Sec:programstructure}
Flow diagram in Fig.~\ref{Victoryflow} demonstrates the {\it victory} implementation of the equations (\ref{GF})-(\ref{DysSchw}). For the interest of the advanced users, we also cross reference to the actual subroutines in the {\em victory} program.
\begin{enumerate}
\item\label{wf:initLambda} Initialize the (fixed) fully irreducible vertex $\Lambda_{r}^{k,k^{\prime}}(q)$, where $r=d/m/s/t$ stands for the different channels. For example, one can set it to the bare interaction as in the parquet approximation (PA) or to the local, fully  irreducible vertex as in the dynamical vertex approximation (D$\Gamma$A). The latter can be obtained by other methods (and program packages), such as exact diagonalization or quantum Monte Carlo.
If $\Lambda_{r}^{k,k^{\prime}}(q)$ is provided in a smaller frequency range than the planned computation range, it is augmented by its asymptotic value (channel dependent constant of the PA).    
\item\label{wf:initF} Initialize $\Sigma_{(0)}(k)$  to  zero or other initial values;  
It is often not of crucial importance to have the initial value close to the actual solution of the problem as {\it victory} iteratively updates the self-energy, but a smarter guess of the initial value will converge the parquet calculations much faster.
At the phase transition boundary convergence may be hard to achieve and it can be helpful to initialize the self-energy with values that effectively start up the calculation far away from the difficult parameter regimes.  
Calculate  ${ G_{(0)}(k)}=(\nu-\epsilon_{\bf k}+\mu-\Sigma_{(0)})^{-1}$ from it; and  initialize $F_{r,(0)}^{k,k^{\prime}}(q)$, $\Gamma_{r,(0)}^{k,k^{\prime}}(q)$ by setting them equal to  $\Lambda_{r}^{k,k^{\prime}}(q)$ or the bare interactions. 
\item\label{wf:redV} Determine the reducible vertex  $\Phi_{r, (n)}^{k,k^{\prime}}(q)$  from the Bethe-Salpeter Eq.~(\ref{PA_F_Phi}). For a fixed value of $q$ these two equations have a standard form of matrix multiplication w.r.t.\ the $k$, and $k^{\prime}$ indices. Therefore, the level-3 BLAS routine {\it zgemm} can be applied straightforwardly. Here, $\Phi_{r, (n)}^{k,k^{\prime}}(q)$ at the $n$th-iteration is calculated from $\Gamma_{r,(n-1)}^{k,k^{\prime}}(q)$ and $F_{r,(n-1)}^{k,k^{\prime}}(q)$ from the \mbox{$(n-1)$th-iteration}. In order to improve convergence in some parameter regime, the calculation of $\Phi_{r, (n)}^{k,k^{\prime}}(q)$ may additionally be damped by its previous solution, i.e. $\Phi_{r, (n),damped}^{k,k^{\prime}}(q)= f_{damping}\cdot \Phi_{r, (n)}^{k,k^{\prime}}(q)+(1-f_{damping})\cdot \Phi_{r, (n-1)}^{k,k^{\prime}}(q)$.
\item\label{wf:kerApp} Due to the structure of the parquet equations, the reducible vertex function is needed in a larger frequency range than what is used in step~\ref{wf:redV}. 
 In~\cite{GangLi-1}, cf.~\cite{Wentzell2016}, it was shown that the high frequency tail of the reducible vertex can be approximated by  the so-called kernel functions  in a sufficient way. Hence in step \ref{wf:ParqEq}., the kernel functions are used  whenever one of the two frequencies $(\nu\:\nu^{\prime})$ exceeds the box $[-N_{\omega},N_{\omega}]$. 
The kernel functions are obtained from the reducible vertex in a level-2 approximation $K^k_{r,2}(q) = \Phi_{r}^{k,(\mathbf{k^{\prime}},N_{\omega})}(q)$ as well as $K^{k^{\prime}}_{r,2}(q) = \Phi_{r}^{(\mathbf{k},N_{\omega}),k^{\prime}}(q)$ in tandem with the constant background obtained in the level-1 approximation $K_{r,1}(q) = \Phi_{r}^{(\mathbf{k},N_{\omega}),(\mathbf{k^{\prime}},N_{\omega})}(q)$.
\item\label{wf:ParqEq} Now every term on the right hand side of the parquet Eq.~(\ref{PA_F}) is known, and a new full vertex function $F_{r, (n)}^{k,k^{\prime}}(q)$ can be obtained. This step is computationally the most involved one due to the size of the two-particle vertex which depends on three momenta and three frequencies and therefore cannot be stored on a single core. Since segments of $\Phi_{r, (n)}^{k,k^{\prime}}(q)$ are stored on different cores and the parquet equations mix different parts of the memory space, extensive intercore communication is required. To this end, we develop an efficient parallelization scheme to broadcast the reducible vertex function across different cores. More details can be found in section~\ref{Sec:parallelization}. 
The new channel-dependent irreducible vertex $\Gamma_{r, (n)}^{k,k^{\prime}}(q)$ is later obtained as $\Gamma_{r,(n)}^{k,k^{\prime}}(q)=F_{r,(n)}^{k,k^{\prime}}(q) - \Phi_{r,(n)}^{k,k^{\prime}}(q)$. 
\item\label{wf:DysSchw} With the full vertex function $F_{r,(n)}^{k,k^{\prime}}(q)$, a new self-energy is then calculated from the Schwinger-Dyson equation~(\ref{DysSchw}).
\item\label{wf:iteration} At this point all terms in equations (\ref{PA_F}) and (\ref{PA_F_Phi}) have been updated once, the calculation can be iterated from step~\ref{wf:redV} to step~\ref{wf:DysSchw} until convergence is achieved. 
\end{enumerate}

\begin{figure}[tb]
\centering
\unitlength=1mm
\linethickness{0.4pt}
\begin{picture}(150.00,132.00)
\put(4.00,5.00){\framebox(146.00,127.00)[cc]{}}
\put(4.00,122.00){\framebox(146.00,10.00)[cc]{
\parbox{14cm}{\ref{wf:initLambda}. initialize ${ \Lambda}^{k,k^{\prime}}_r(q)$ with $r\in (d,s,m,t)$    } } }
\put(110.00,125.00){[{\em readin}]} 
\put(4.00,105.00){\framebox(146.00,17.00)[cc]{
\parbox{14cm}{
\ref{wf:initF}. initialize ${\Sigma}(k)$ and obtain ${ G(k)}=(\nu-\epsilon_{\bf k}+\mu-\Sigma)^{-1}$ 
\vskip 0.2cm \hskip 0.07cm
\phantom{\ref{wf:initF}.} initialize ${ F}^{k,k^{\prime}}_r(q) = { \Gamma}^{k,k^{\prime}}_r(q) = \Lambda^{k,k^{\prime}}_r(q)$ 
}}}
\put(110.00,116.00){[{\em pa\_Gkw\_Chi0}]}
\put(110.00,109.00){[{\em readin}]}

\put(7.00,11.00){
\parbox{12.5cm}{
\ref{wf:iteration}. iterate with  $ { \Sigma} \!=\! {\Sigma}_{\rm new}$ until convergence, i.e.\ $ ||{ \Sigma} - { \Sigma_{\rm new}}|| < \varepsilon $.
}}
\put(110.00,97.00){[{\em reducible\_vertex}]}
\put(110.00,81.50){[{\em get\_kernel\_function}]}

\put(12.00,75.00){\framebox(138.00,30.00)[cc]
 {\parbox{13.3cm}{
\ref{wf:redV}. calculate the reducible vertex ${ \Phi_r}$ 
\begin{align*}
\Phi_r \stackrel{Eq.~(\ref{PA_F_Phi})}{=} \Gamma_r \circ G G \circ F_{r}
\end{align*}
\ref{wf:kerApp}. determine the kernel functions ${K}_1(q)$, ${K}^{k}_2(q)$ and ${K}^{k'}_2(q)$ 
}}}
\put(110.00,68.00){[{\em solve\_parquet\_equation}]}
\put(12.00,45.00){\framebox(138.00,30.00)[cc]
{\parbox{13.3cm}{
\ref{wf:ParqEq}. calculate the complete vertex ${ F_r}$ 
\begin{align*}
F_r \stackrel{Eq.~(\ref{PA_F})}{=} \Lambda_r + c_i\Phi_i\quad \text{for } i\in (d,s,m,t)
\end{align*}
\quad and update the channel-dependent irreducible vertex ${ \Gamma_r} = F_r - \Phi_r$ 
}}}

\put(110.00,37.00){[{\em self\_energy}]}
\put(110.00,26.00){[{\em pa\_Gkw\_Chi0}]}

\put(12.00,20.00){\framebox(138.00,25.00)[cc]
{\parbox{13.3cm}{
\ref{wf:DysSchw}. calculate the self-energy ${\Sigma}$ 
\[
\Sigma \stackrel{Eq.~(\ref{DysSchw})}{=} U\circ GGG \circ F
\]
\quad and update the Green's function ${ G(k)}=(\nu-\epsilon_{\bf k}+\mu-\Sigma)^{-1}$ 
}}}
\end{picture}
\caption{Flow diagram for solving the parquet equations in {\it victory}. The name of the subroutine used at each step is provided in the rectangular brackets.}
\label{Victoryflow}\label{flowVictory}
\end{figure}

\subsection{Technical remarks}
\label{Sec:technical}
\begin{itemize}
\item  In {\it victory}, each vertex function is implemented as a two-dimensional array for a given value of $q$. Here, $k, k^{\prime}$ and $q$ are momentum-frequency mixed variables where each of them can be referred to in the program by a unique integer index. There are two types of indexing in {\it victory}. A combined momentum-frequency index stored as an integer in the range $[1,2\!\cdot\! N_{\omega}\!\cdot\! N_c]$, with $N_c$ being the momentum cluster size and $2\!\cdot\! N_{\omega}$ the number of frequencies; and a derived-type index map containing integers which directly refer to components of momentum ${\mathbf k}$ and Matsubara frequency $\nu$ (or in case of a bosonic frequency to $\omega$).
The three linear algebra operations needed in Eq.~(\ref{PA_F}), {\it i.e.} $k+k^{\prime}+q$, $k^{\prime}-k$ and $q-k-k^{\prime}$, can thus be easily defined by using the derived-type index maps [subroutine \mbox{{\em index\_operation}}]. There is a unique mapping between these two types of indexing [subroutine \mbox{{\em list\_index}}].  
\item The memory consumption of the program is proportional to the cube of the range of each argument, {\it i.e.} $k$, $k^{\prime}$ and $q$. The number of momentum points is simply determined by the size of the cluster on which the calculations are performed. The number of Matsubara frequencies is given by the user as input, but it is essentially determined by the temperature of the system, {\it i.e.} the lower the temperature the more Matsubara frequencies are required. The memory consumption for three different momentum cluster sizes is given in Table~\ref{MemoryConsumption}. 
\begin{table}[tb]
\centering
\begin{tabular}{||c|c|r|r||}
\hline
Cluster size & No. of cores & Memory/core & Total memory used\\
\hline\hline
$2\times2$ & $32$ & $0.09$ GB & $2.9$ GB \\
$4\times4$ & $128$ & $1.7$ GB & $217.6$ GB \\
$6\times6$ & $288$ & $14.7$ GB & $4.2$ TB \\
\hline
\end{tabular}
\caption{Memory consumption for different momentum cluster sizes. The data are for $N_{\omega}=32$. The segments of vertex functions are distibuted such that each core stores data for $4$ values of the combined momentum and frequency index $q$ (cf. Fig.~\ref{fig:parallelIllust}). Copies of functions other than vertex (e.g. Green function, self-energy, kernels) are stored on each core to avoid extensive communication. This causes the actual total memory consumption to grow slightly faster than the cube of the cluster size (with a prefactor depending on the number of cores used). }
\label{MemoryConsumption}
\end{table}

\item A practical problem arises due to the finite size of the frequency box. As a result of linear operations of the arguments mentioned before, {\it i.e.}   $k+k^{\prime}+q$, $k^{\prime}-k$ and $q-k-k^{\prime}$,  we stay in a bigger parameter space than the one defined individually for $k$, $k^{\prime}$ and $q$. This fact would in the end make the loop iteration impossible as the evaluation of $F_{r}^{k,k^{\prime}}(q)$ in the parquet Eq.~(\ref{PA_F}) requires $\Phi_{r}$ in a bigger parameter space. 
As a solution we developed the two-level kernel approximation which is discussed in step~\ref{wf:kerApp} of the program structure and explained in detail in our previous works~\cite{GangLi-1, Wentzell2016}. The kernel approximation respects the correct asymptotics of the reducible vertex function $\Phi_{r}$ and extends the calculation of Eq.~(\ref{PA_F}) to arbitrarily large parameter space.
\item The limitation to a finite frequency range  without  periodicity conditions, e.g. as the ones for the momentum, affects not only the calculation of the full vertex functions $F$ but also the determination of the self-energy in the Schwinger-Dyson Eq.~(\ref{DysSchw}). In general, the truncation of the two sums seriously influences the high-frequency tails of $\Sigma$. However, with the help of the kernel approximations these sums can be regulated, leading to a correct behavior of the self-energy. That is,  we take the bare interaction as $F$ in Eq.~(\ref{DysSchw}) for arbitrary large frequencies and the kernel functions for a ({\em e.g.} two or three times) larger frequency box. 
 A more detailed discussion thereof is likewise provided in \cite{GangLi-1}.
\item Whenever possible, we evaluate the lowest order diagram (the "bubble" diagram of the Green's function) by using the Fast-Fourier-Transform between the Matsubara frequency $\nu$ and imaginary-time $\tau$. The high-frequency asymptotic tail of the bubble diagram is explicitly evaluated. This scheme is used in the evaluation of the reducible vertex function $\Phi_{r}^{k,k^{\prime}}(q)$ to substitute the discrete frequency sum of the bubble part.
\end{itemize}
  
\subsection{Parallelization}
\label{Sec:parallelization}
\begin{figure}[htbp]
\centering
\includegraphics[width=.5\linewidth]{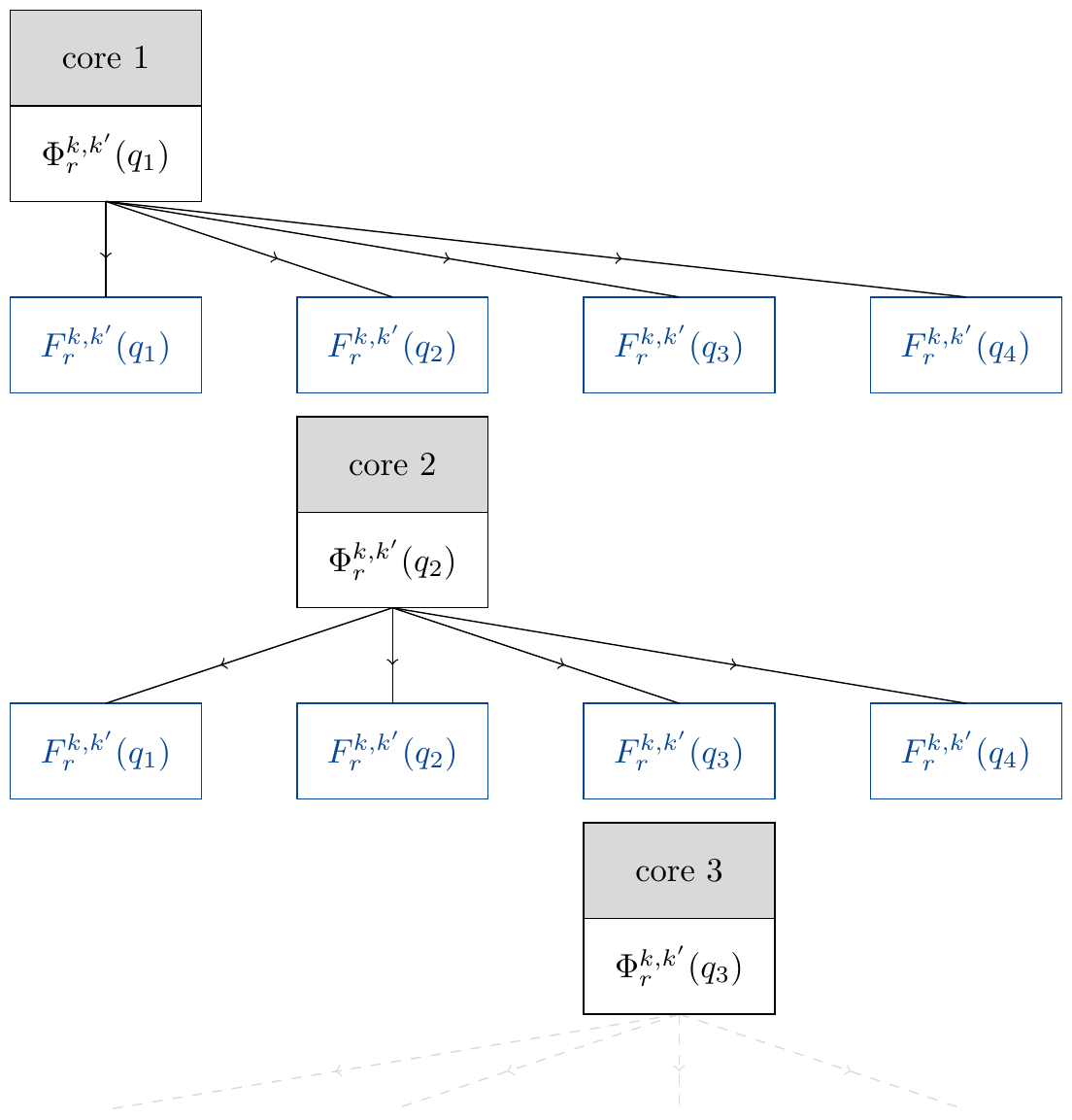}
\caption{The data for the vertex functions is distributed with respect to the  combined momentum and bosonic frequency index $q$ on different cores (an example with 4 cores is illustrated). This results in necessary all-to-all communication among the cores when calculating the full vertex function $F$ in Eq.~(\ref{PA_F}). For example, updating the vertex $F$ at ($k$, $k^{\prime}$, $q_2$) on core 2 requires among others information from  $\Phi$ at ($k$, $k+k^{\prime}$, $q_1=k^{\prime}-k$) which is, say, stored on core 1. The other combinations of $k^{\prime}$ and $k$ require information from all other cores. Likewise the $\Phi$  of core 2 is needed by all other cores and hence has to be broadcast to all cores.}
\label{fig:parallelIllust}
\end{figure}
In the current release of {\it victory}, we provide parallelization based on Message Passing Interface (MPI). The central issue in the parallelization is the memory consumption, since the most memory-consuming objects in the program, {\it i.e.} the vertex functions, cannot be stored in the memory of one core only. They are distributed over cores according to their $q$-values, for illustration see Fig~\ref{fig:parallelIllust}.
Correspondingly, each core carries different segments of the vertex functions. 
The Bethe-Salpeter equation~(\ref{PA_F_Phi}) is not affected by the parallelization as, for a fixed value of $q$, these equations can be evaluated at each core without communication with others.

However, the calculation of the parquet equation~(\ref{PA_F}) requires data exchange among all cores since different frequencies and momenta are mixed. 
In other words, each core has to send its data to all other cores.
At the same time it also receives data from all other cores, which essentially requires a (possibly) non-blocking communication among different cores. 
This communication becomes the actual bottleneck of the calcuation.

In {\it victory} we have tested two flavors of parallelization: (1) One can combine a non-blocking \mbox{MPI\_ISEND} with a blocking \mbox{MPI\_RECV} to ensure that every core receives the required data before starting to solve the parquet equation.  (2)  A more economical way is to use \mbox{MPI\_BCAST} {\em i.e.} each core distributes its segment of vertex function to all other cores. The advantage of using collective communication in this case is in the decreased number of actual point-to-point communications (for \mbox{MPI\_BCAST} with a binomial tree algorithm it is $N_{cores}\!\cdot\!\log N_{cores}$ as compared to the $N_{cores}^2$ needed by \mbox{MPI\_ISEND} and \mbox{MPI\_RECV}). This outweighs the potential gain in time that the non-blocking communication provides. Therefore, in the current version of {\it victory} only collective (blocking) communication is implemented. 

Another parallelization execution occurs when the self-energy function is calculated in step~\ref{wf:ParqEq}. 
After the evaluation of the parquet equation, the full vertex function $F^{k,k^{\prime}}_{r}(q)$ is updated at all cores. At each core only a segment of $F^{k,k^{\prime}}_{r}(q)$ is stored, as explained before.
The calculation of the self-energy requires however $F^{k,k^{\prime}}_{r}(q)$ at all $q$ values, cf. Eq.~(\ref{DysSchw}). Hence further communications are needed. 
Based on the parallelization strategy we established, we perform a partial sum over the $q$-values available at each core. Each core stores now only part of the self-energy and the sum over all these parts is the correct self-energy.  
The results from all cores are collected with \mbox{MPI\_ALLREDUCE}, which yields the desired self-energy function at each core. 

We note that, in {\it victory}, only the two-particle vertex functions are distributed over cores, all other quantities such as the single-particle Green's function, self-energy, as well as the two-level kernel functions are stored on all cores to reduce the demand on core communications. 

\section{Example case}
\label{Sec:example}
To illustrate the capability of {\it victory}, we consider here a single-band Hubbard  model on a square lattice. The results presented in this section correspond mainly to a $4\times 4$ cluster with periodic boundary conditions. 
For the purpose of comparison, in some cases we also show the results for a $6\times6$ cluster. 
As stated before, the parquet equations guarantee self-consistency at both single- and two-particle levels. Correspondingly, in {\it victory} one gets both the single- and two-particle Green's function from the same calculation.  
All the results presented in this section were obtained using the parquet approximation (PA).

\subsection{Single-particle quantities}
\label{Sec:singleparticle}
 
\begin{figure}[htbp]
\centering
\includegraphics[width=\linewidth]{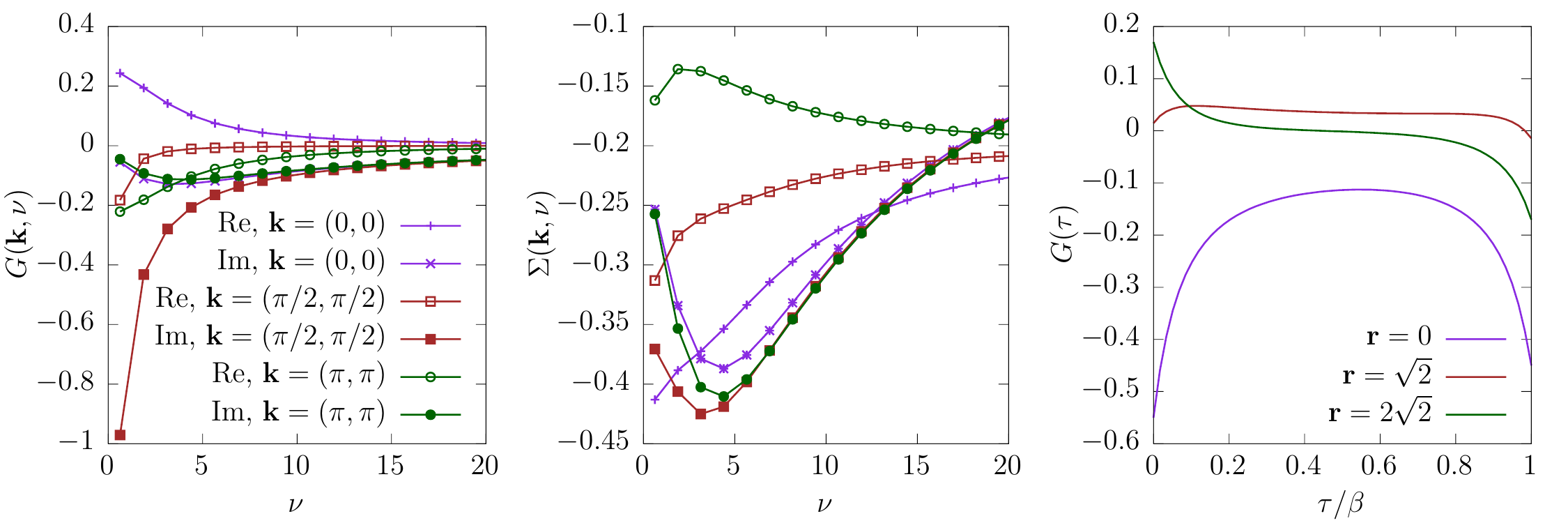}
\caption{Real and imaginary part of the single-particle (a) Green's function and (b) self-energy as a function of Matsubara frequency $\nu$ at three different cluster momenta. (c) shows the imaginary-time Green's function in real-space at different distances $r$. The parameters for the 2D Hubbard model are $\beta t= 5$, $U/t = 4.0$ (here and in the following $t\equiv 1$ sets our unit of energy) and average particle density $\langle n\rangle=0.9$.}
\label{Fig:single}
\end{figure}
In a finite-size cluster, one has only a limited number of independent cluster momenta at which the single-particle Green's functions behave differently. In {\it victory}, we calculate the single-particle Green's function and two-particle vertex functions at all cluster momenta as it is straightforward to manipulate the internal momentum sum in the full Brillouin zone. However, we stress that this is not really necessary, {\it i.e.} a calculation with only the independent cluster momenta of the irreducible Brilluin zone can also be conducted after some further modifications of {\it victory}.    
In Fig.~\ref{Fig:single}(a), the imaginary part of the single-particle Green's function for the 2D Hubbard model at different independent cluster momenta for the $4\times4$ cluster is shown as a function of Matsubara frequency $\nu$ for $\langle n\rangle=0.9$. The current implementation of {\it victory} works equally for undoped and doped cases. 
The imaginary part of the Green's function in Fig.~\ref{Fig:single}(a) displays a clear momentum selectivity, in particular at ${\bf k} = (\pi/2, \pi/2)$ it quickly increases in amplitude and develops a clear metallic character (corresponding to a $1/\nu$  shape at small $\nu$). In contrast, at ${\bf k}=(0, 0)$ and $(\pi,\pi)$, the Green's functions approach zero with the decrease of frequency, suggesting an insulating state at these momenta.

As $U/t=4$ is still in the weak-correlation regime for the 2D Hubbard model ({\it i.e.} it is smaller than the critical interaction strength for the metal-insulator transition), the system essentially follows the same character as  in the non-interacting limit.  That is,
${\bf k}=(\pi/2, \pi/2)$ is on the Fermi surface, while the other two momenta are far away from it. 
The self-energy, as a measure of the effective correlations, is shown in Fig.~\ref{Fig:single}(b). 
As one generally expects, the electronic correlations renormalize states at the Fermi surface more significantly than  the states with higher binding energies. 
Consequently, the self-energy at ${\bf k}=(\pi/2, \pi/2)$ is somewhat bigger than at the other two momenta presented.

Fig.~\ref{Fig:single}(c) displays the real-space Green's function as a function of the imaginary-time.  Here, $r=0$ corresponds to the local Green's function, $r=\sqrt{2}$ and  $r=2\sqrt{2}$ to the nonlocal one between site (0,0) and
(1,1) and (2,2), respectively.
In {\it victory}, the average particle number $\langle n\rangle$ is self-consistently determined by adjusting the chemical potential $\mu$ to satisfy $\langle n\rangle=\sum_{\sigma}G_{r=0,\sigma}(\tau=0^{-})$.

\begin{figure}[htbp]
\centering
\includegraphics[width=\linewidth]{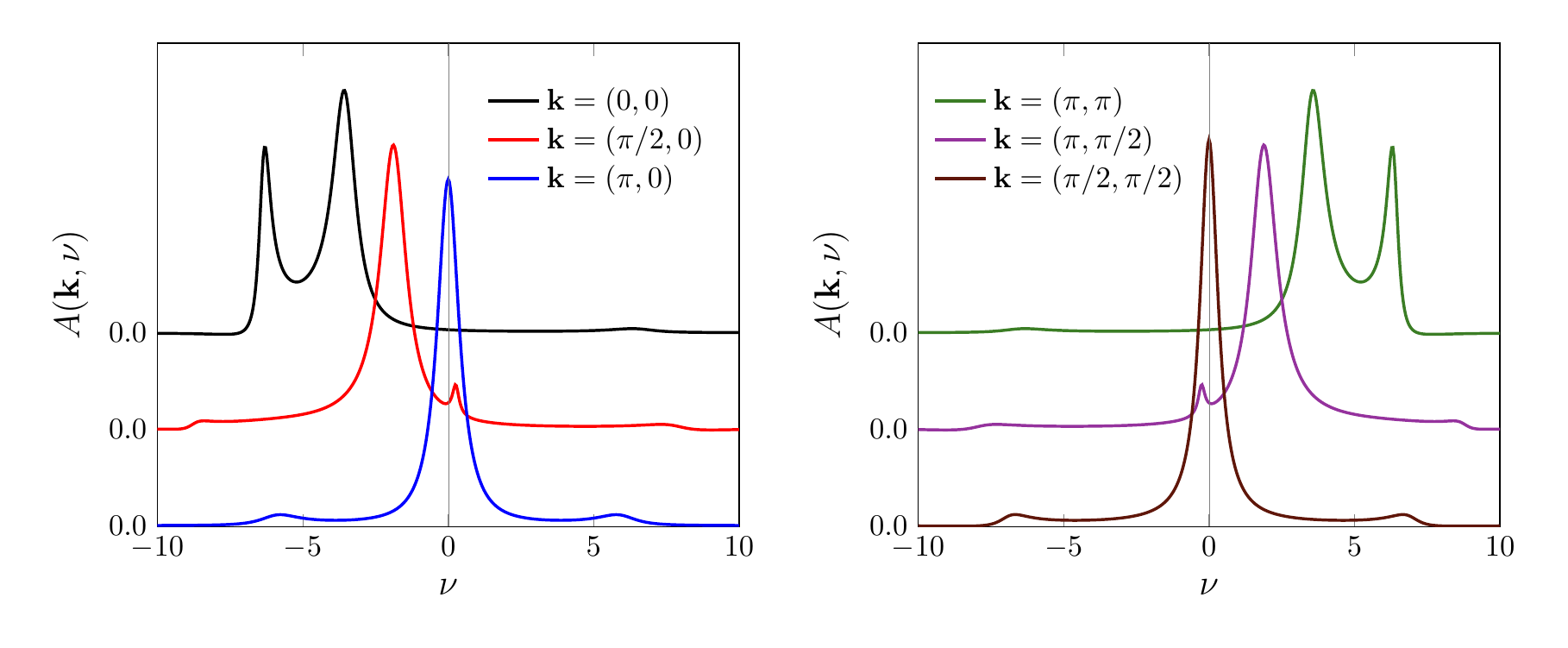}
\caption{Spectral function for different momenta $\bf{k}$. The parameters are $\beta t = 2$, $U/t = 4.0$ and the average particle density is $\langle n\rangle=1.0$. The grey vertical line denotes the Fermi energy.}
\label{Fig:spectral}
\end{figure}

Obtaining quantities on the real frequency axis is not implemented directly in  {\it victory}. The Matsubara Green's function can however be used as input into a Pad\'e analytic continuation script (not provided) to obtain momentum dependent spectral functions as presented in Fig.~\ref{Fig:spectral}. Other analytical continuations are of course also possible.  
The results were obtained for the following parameters $\beta t = 2$, $U/t = 4.0$ and  $\langle n\rangle=1.0$ for a $4\times4$ momentum cluster. A clear quasi-particle peak is visible at momenta on the Fermi surface: ${\bf k}=(\pi,0)$ and $(\pi/2, \pi/2)$. As we move away from the Fermi surface, there is also a shift of some spectral weight to higher (lower) frequencies. For $U/t = 4.0$,  only for these momenta away from the Fermi surface there is  a two peak structure corresponding to the quasiparticle peak and the upper (lower) Hubbard band. 

\subsection{Two-particle quantities}
\label{Sec:twoparticle}
A major output of {\it victory} are various two-particle quantities, which include the two-particle full vertex function $F_{r}(k,k^{\prime};q)$, the channel-dependent reducible vertex functions $\Phi_{r}(k,k^{\prime};q)$ and the channel-dependent irreducible vertex functions $\Gamma_{r}(k,k^{\prime};q)$. 
These vertex functions play important roles in many-body theory. For example, the magnetic, charge and pairing response functions are readily obtained by attaching four legs to the full vertex function $F_{r}(k,k^{\prime};q)$.
This yields, 
 after summing over $k$ and $k^{\prime}$ and adding the bubble contribution, the physical responses which are directly observable in corresponding experiments, such as neutron scattering. 

\begin{figure}[htbp]
\centering
\includegraphics[width=\linewidth]{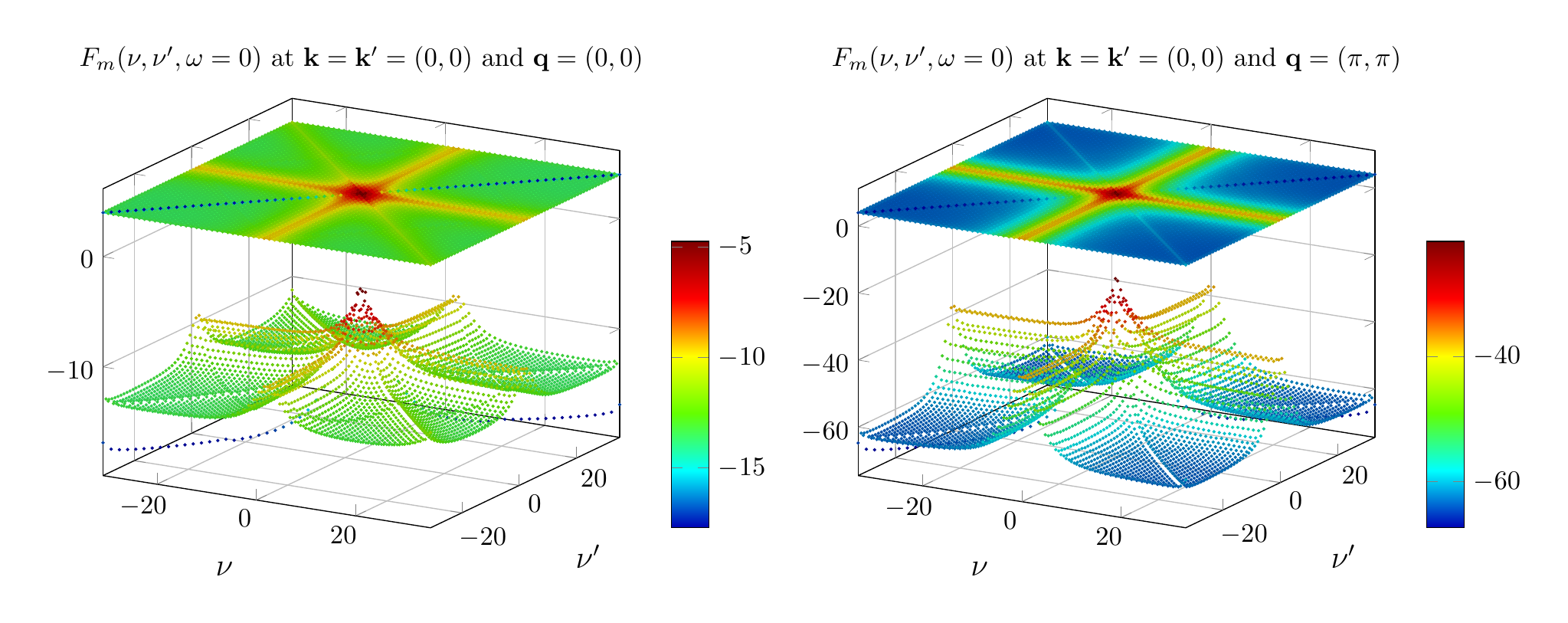}
\includegraphics[width=\linewidth]{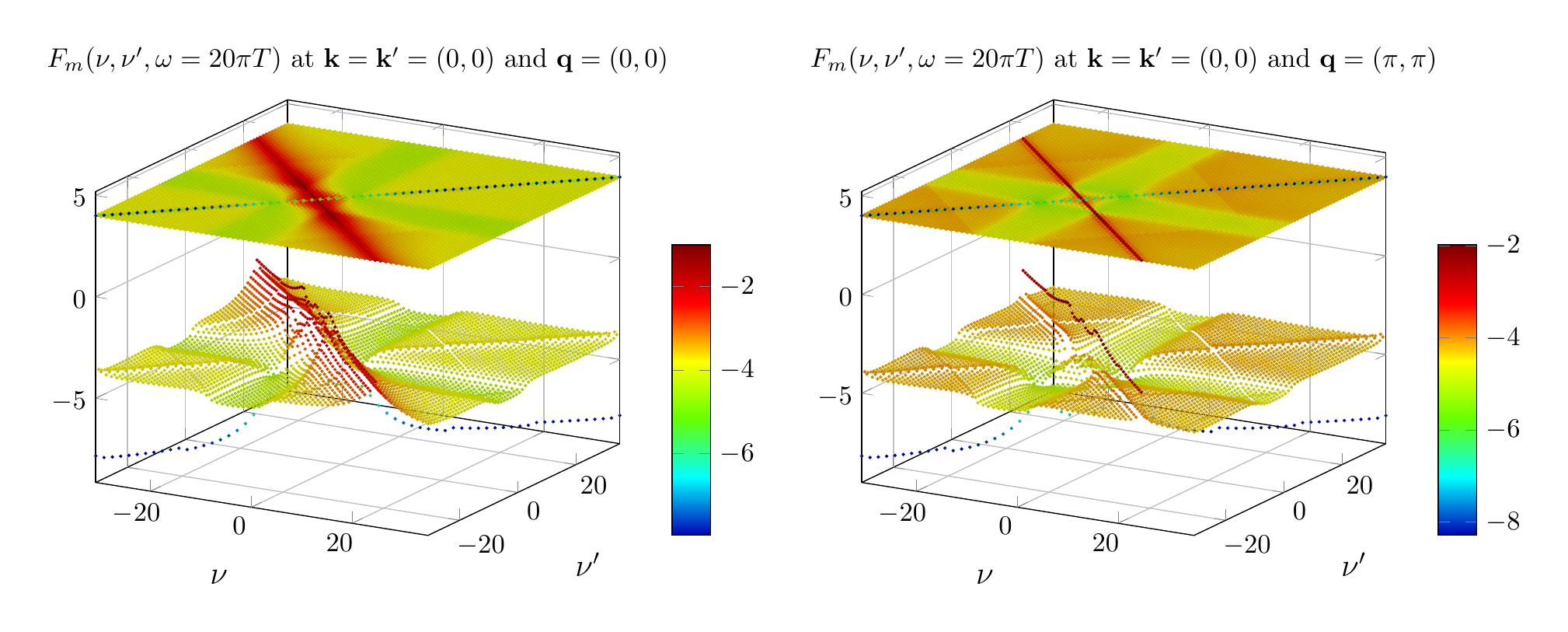}
\caption{Two-particle full vertex function $F_{m}(k,k^{\prime};q)$ as a function of Matsubara frequencies at two different transfer momenta  ${\bf q}=(0, 0)$ (left) and ${\bf q}=(\pi,\pi)$ (right)  for $\beta t = 6$, $U/t = 4.0$, and $\langle n\rangle=1.0$.  The plots in the first and the second row correspond to transfer frequencies $\omega=0$ and $\omega=20\pi T$, respectively. Each panel includes two representations of the same vertex:  a two dimensional intensity plot with the amplitude of the vertex given by the color bar and a three dimensional plot with both $z$-axis and color giving  the amplitude of the vertex.}
\label{Fig:Fm}
\end{figure}

In Fig.~\ref{Fig:Fm}, the full vertex functions in the magnetic channel are shown as functions of the two fermionic Matsubara frequencies $\nu$ and $\nu^{\prime}$. The transfer frequencies are taken as $\omega=0$ and $\omega=20\pi T$ for the first and the second row, respectively, and the parquet equations have been solved on a $6\times 6$ cluster in momentum space.  
To illustrate the momentum dependence of the vertex function, we have chosen different transfer cluster momenta, {\it i.e.}  ${\bf q}=(0, 0)$ and ${\bf q}=(\pi,\pi)$ at ${\bf k}={\bf k}^{\prime}=(0, 0)$. 

On the square lattice, the single-band Hubbard model shows strong antiferromagnetic spin fluctuations  at ${\bf q}=(\pi,\pi)$. This is reflected 
in the  strongly peaked structure of the two-particle full vertex function $F_{m}$ for ${\bf q}=(\pi,\pi)$  in Fig.~\ref{Fig:Fm}.
The frequency structure of the vertex functions is better seen in the intensity plots shown at the top of each vertex figure. 
As for transfer frequency $\omega=0$, the major vertex structure is the cross appearing at $\nu=0$ and $\nu^{\prime}=0$. 
The cross shifts to $\nu=-\omega$ and  $\nu^{\prime}=-\omega$ when $\omega$ becomes nonzero. At the same time, the diagonal structure at $\nu+\nu^{\prime}+\omega=0$ becomes more prominent.
This can be clearly seen by comparing the second row of Fig.~\ref{Fig:Fm} to the first row,  and it essentially originates from the special momentum and frequency dependence of the parquet equation~(\ref{PA_F}).  
\begin{figure}[htbp]
\centering
\includegraphics[width=\linewidth]{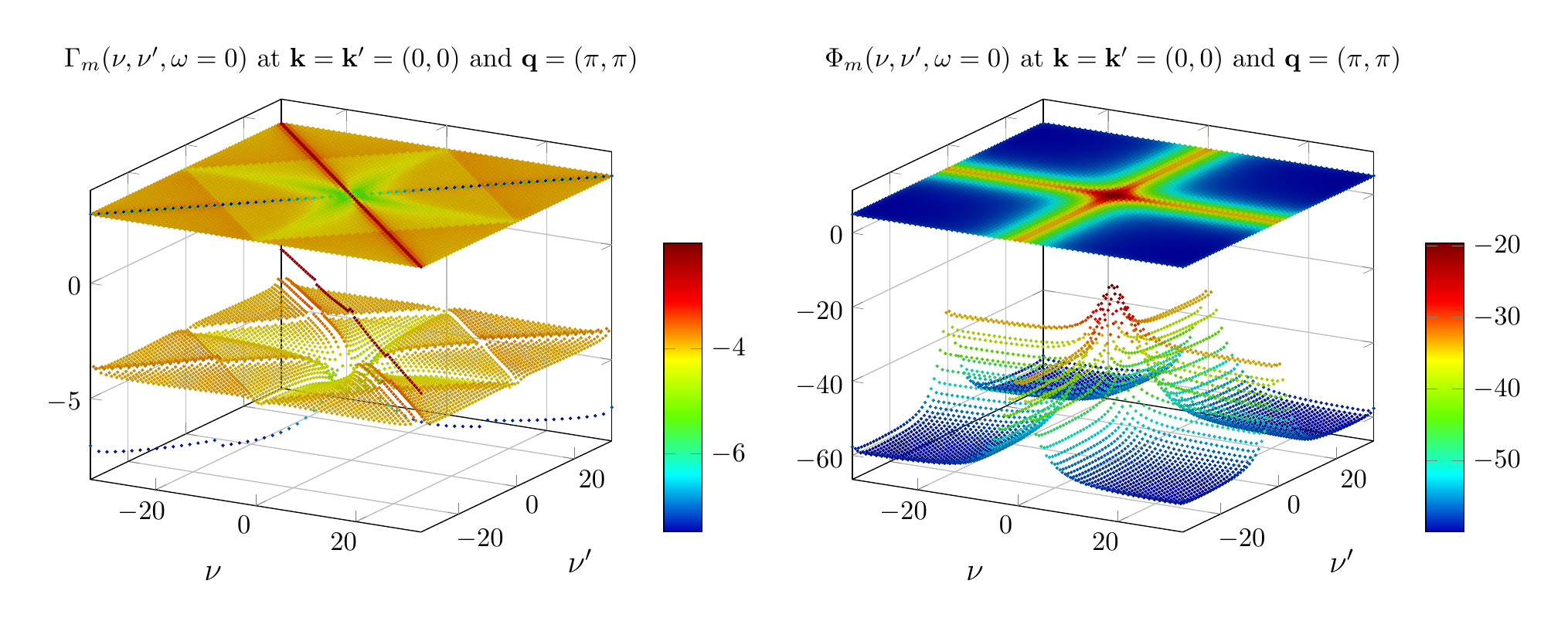}
\caption{The two-particle irreducible vertex function $\Gamma_{m}(k,k^{\prime};q)$ and the reducible vertex function  $\Phi_{m}(k,k^{\prime};q)$ as a function of Matsubara frequencies at transfer frequency $\omega=0$ and transfer momentum ${\bf q}=(\pi,\pi)$ in a $6\times6$ cluster. The parameters are the same as in Fig.~\ref{Fig:Fm}.}
\label{Fig:Gm}
\end{figure}

The full vertex $F$ of Fig.~\ref{Fig:Fm} can be separated, in each channel $r$, into its irreducible and reducible components $F_r=\Gamma_r+\Phi_r$.
Fig.~\ref{Fig:Gm} shows both of these components for the magnetic channel and 
 the biggest contribution at  ${\bf q}=(\pi,\pi)$, $\omega=0$.
This reveals that the large cross-like contribution to $F_m$ stems from the reducible 
vertex $\Phi_m$. That is, it is built up through the Bethe-Salpeter ladder in the magnetic channel, as is to be expected for strong antiferromagnetic spin fluctuations. Some other contributions such as the $\nu=\nu^{\prime}$ diagonal structure that is a bit off-set from the rest of the vertex are contained in $\Gamma_m$. Indeed we know from perturbation theory that this diagonal stems from the particle-hole transversal channel \cite{RMPVertex,PhysRevB.86.125114} which is irreducible in the magnetic particle-hole channel shown in Fig.~\ref{Fig:Gm}.

The effect of using a finite frequency box is clearly visible in the irreducible vertex $\Gamma_m$ in the left panel of Fig.~\ref{Fig:Gm}. It manifests itself as a square-shaped structure at larger fermionic frequencies. It is also visible, though with smaller relative intensity, in the full vertex function $F_m$ shown in Fig.~\ref{Fig:Fm}. This finite-size effect disappears when a large enough frequency range is taken. For the data presented in Figs.~\ref{Fig:Fm} and~\ref{Fig:Gm} we used $N_{\omega}=32$.   

We want to note that by using {\it victory} one can also obtain various susceptibilities  after further manipulating the vertex data. We do not provide such scripts in the standard {\it victory} package as they can be easily implemented by the users. 
Furthermore, the vertex functions calculated in {\it victory} can also serve as the input for various many-body methods. 
For example, the dual-fermion approach~\cite{Rubtsov2008, PhysRevLett.102.206401} and the non-local expansion scheme~\cite{Li2015} both use the local full vertex function to construct the non-local self-energy diagram;
the two-particle irreducible vertex $\Gamma_{r}$  is the building block in the ladder flavor of D$\Gamma$A~\cite{Toschi2007, doi:10.1143/PTPS.176.117} and the $1$-particle irreducible approach~\cite{Rohringer2013}. 
In this respect,  the {\it victory} package not only provides a practical implementation of the parquet method but also serves as a calculator of various vertex functions, for which a reliable numerical scheme is still missing (in particular with the momentum resolution). 
It thus can be combined with other many-body methods to study more complicated electronic systems.

\begin{figure}[htbp]
\centering
\includegraphics[width=\linewidth]{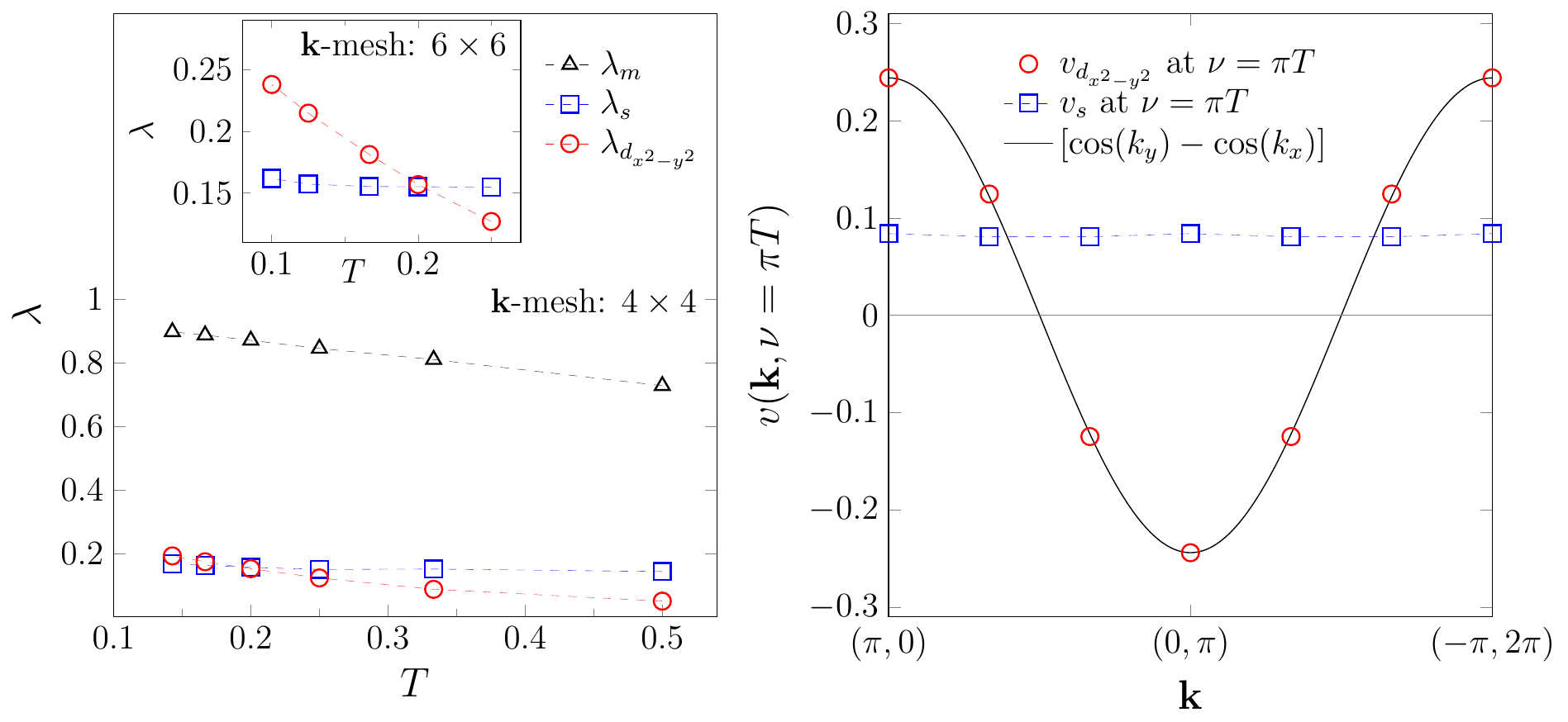}
\caption{Left: The temperature dependence of the leading eigenvalues of the Bethe-Salpeter equation  in the magnetic channel $\lambda_{m}$ and in the particle-particle channel $\lambda_{s}$ and $\lambda_{d_{x^{2}-y^{2}}}$ for the Hubbard model with $U = 4t$ and half filling  $\langle n\rangle = 1$. Here, the subscript  $s$ and $d_{x^{2}-y^{2}}$ denotes the symmetry of the corresponding eigenvectors $v_{s}$ and $v_{d_{x^{2}-y^{2}}}$ 
whose momentum dependence  is shown in the right panel for the $6\times6$ cluster and  the first Matsubara frequency at $\beta t =6$.
The dominant eigenvalue is the magnetic one, however, with decreasing temperature the $d$-wave eigenvalue becomes larger than that for the $s$-wave  in the particle-particle channel. The inset shows the  $6\times6$ cluster in comparison to the  $4\times4$ cluster in the main left panel.}
\label{Fig:lambda}
\end{figure}

Another interesting quantity that is examined in {\it victory} are the eigenvalues of the Bethe-Salpeter equation in all four above-mentioned channels. These can be obtained as solutions of the following eigen-equations:
\begin{subequations}
\begin{align}
-\frac{T}{N}\sum_{k^{\prime}}\Gamma_{d/m}(k, k^{\prime};q)G(k^{\prime})G(k^{\prime}+q) v_{q}(k^{\prime}) = \lambda_q(k) v_{q}(k)\;,\\
\pm\frac{T}{2N}\sum_{k^{\prime}}\Gamma_{t/s}(k, k^{\prime};q)G(k^{\prime})G(q-k^{\prime})v_{q}(k^{\prime})=\lambda_q(k) v_{q}(k)\;.
\label{Eq:EV}
\end{align}
\end{subequations} 
The corresponding eigenvalue,
$\lambda_q(k)$, provides a simple measure of the stability of the parquet equations.
 The Bethe-Salpeter equation, by iteratively appending the irreducible vertex $\Gamma_{r}$ to the corresponding full vertex $F_{r}$, determines the ladder contribution to the corresponding two-particle fluctuations in the given channel. 
 The {\it victory} parquet formalism is based on the translational symmetry, SU(2) symmetry and time-reversal symmetry, a divergence of the Bethe-Salpeter equation tells us that one of these symmetries is broken and we approach a phase transition.  A phase transition occurs, more specifically, when the largest eigenvalue for transfer frequency $\omega = 0$ approches 1. Note that we cannot calculate the vertex function in a symmetry-broken phase in the current implementation, albeit this is in principle possible.

In {\it victory}, the eigenvalue $\lambda_q(k)$ is given at different transfer momenta ${\bf q}$. 
The largest $\lambda$ typically appears at values of ${\bf q}$ commensurate to the lattice periodicity. 
For example, on a square lattice one expects $\lambda$ at ${\bf q}=(\pi, \pi)$ in the magnetic channel to be most relevant. 
In the case of a triangular lattice, on the other hand, it is at ${\bf q}=(2\pi/3, 2\pi/3)$. 
For a possible charge density wave instability, e.g. in the presence of non-local Coulomb interactions, the leading eigenvalue in the charge channel is of particular interest.  

In Fig.~\ref{Fig:lambda} we show the largest eigenvalue from the magnetic and, more interestingly,  the pairing channel where the paring vertex $\Gamma_{\overline{\uparrow\downarrow}}$ is obtained from the linear combination of the corresponding singlet and triplet components,
$\Gamma_{\overline{\uparrow\downarrow}}=\frac{1}{2}(\Gamma_{s} - \Gamma_{t})$.  

On the square lattice, the single-band Hubbard model at half-filling is unstable against antiferromagnetic ordering at arbitrary interaction at zero temperature. Concomitant with this,  Fig.~\ref{Fig:lambda} shows a monotonous increase of $\lambda$ in the magnetic channel which approaches 1 at low temperatures. 
In the paring channel, we show the leading eigenvalues corresponding to the $s$-wave and $d$-wave paring symmetries, respectively. 
The paring symmetry can be easily identified by inspecting the symmetry of the corresponding wave function $v_{q}(k)$  
of Eq.~(\ref{Eq:EV}) in momentum space. It is as shown in the right panel of Fig.~\ref{Fig:lambda} for $s$- and $d$-wave pairing symmetry. 
At high temperatures $s$-wave is dominating and gradually gives way to the $d_{x^{2}-y^{2}}$ wave at low temperatures. 
This is seen more clearly in the $6 \times 6$ cluster in the inset of the left plot in Fig.~\ref{Fig:lambda}. At half-filling, however, the antiferromagnetic eigenvalue  $\lambda_m$ still prevails (is closer to the divergence at $\lambda_m=1$).
The eigenvalues and eigenfunctions at each channel are a direct output of {\it victory}.

\section{Conclusions and Outlooks}
\label{Sec:conclusion}
In this work, we presented a Fortran implementation of the parquet equations for the single-band Hubbard model.  For a given fully irreducible vertex $\Lambda$, the parquet equations allow us to calculate all one- and two-particle correlation functions. The computational bottleneck is the memory consumption since the vertices are large objects that depend on three  frequencies and momenta.

We have implemented two-level kernel functions to correctly account for the asymptotic frequency behavior of the two-particle vertex function in Matsubara frequency space. Employing this asymptotic behavior is essential to do calculations with a finite frequency box. 

 The program is parallelized using mostly collective communication modes (one-to-all and all-to-all) which significantly decreases the number of point-to-point communications.
Without further optimization, the current program is already feasible for the study of  the Hubbard model on a $6\times6$ cluster for an arbitrary number of electrons.

Further improvements of this package to include non-local Coulomb interactions and to reduce the number of cluster momenta in the calculations by employing the point group symmetry are currently under development.

\section{Acknowledgment}
We are grateful for valuable discussions with Partrick Tunstr\"{o}m, Nils Wentzell, Agello Valli, Angese Tagliavini, Georg Rohringer, Alessandro Toschi, Sabine Andergassen. We also greatly acknowledge the help of Claudia Blaas-Schenner in optimizing the MPI parallelization scheme.   
GL acknowledges the financial support from the starting grant of ShanghaiTech University.  AK and KH have been supported  by the European Research Council under the European Union's Seventh
Framework Program (FP/2007-2013) through  ERC grant agreement n.\ 306447;
Part of the  computational  results presented has been obtained using the VSC. 

\bibliographystyle{elsarticle-num} 
\bibliography{ref}

\begin{thebibliography}{10}
\bibitem{footnote1}
That is, cutting a Green's function line does not separate the Feynman diagram
  into two pieces.
\end{thebibliography}

\end{document}